\documentclass{webofc}

\usepackage[varg]{txfonts}   
\usepackage{hyperref}
\usepackage{url}
\usepackage[
    backend=biber,
style=numeric-comp,
sorting=none
]{biblatex}
\RequirePackage{orcidlink}

\hypersetup{colorlinks=true,citecolor=blue,urlcolor=blue,linkcolor=blue}
\begin{document}
\title{A Phase-Space Inclusive Figure of Merit Based in Optimal Transport for Validating Monte Carlo Reweightings}
%
%

\author{
        \firstname{Rishabh} \lastname{Jain}\inst{1}\orcidlink{0009-0004-5004-7357}\fnsep\thanks{\email{rishabh_jain@brown.edu}} \and \firstname{Lauren} \lastname{Hay} \inst{1,2}\orcidlink{0000-0002-7086-7641} \and
        \firstname{Matt} \lastname{LeBlanc} \inst{1,2}\orcidlink{0000-0001-5977-6418} \and
        \firstname{Jennifer} \lastname{Roloff} \inst{1}\orcidlink{0000-0001-6479-3079}
}

\institute{Department of Physics, Brown University, Providence, RI, USA
\and
           The NSF AI Institute for Artificial Intelligence \& Fundamental Interactions, Boston, MA, USA}

\abstract{Validating whether the underlying physics of a model has been retained after a full phase-space reweighting poses a unique challenge. Often, validation relies on comparing histograms of 1D observables; however, this can mask correlations and biases in the complete prediction. We present a novel, unbinned approach to comparing the performance of such reweighting schemes based on the ``Cross-Section-Mover’s Distance'', an application of Optimal Transport that quantifies the work required to transform one theoretical prediction into another and enables an interpretation of results in terms of metric spaces. We demonstrate its utility when benchmarking various reweighting schemes that mitigate the effects of negative weights in a Monte Carlo simulation. This approach can be broadly applied in other scenarios where biases in full phase-space reweighting schemes should be studied in an unbinned way.
}
\maketitle
\section{Introduction}
\label{intro}
Comparing the compatibility of two datasets occurs under many different contexts such as unfolding, generative modeling, negative weight mitigation, and more. Due to the high dimensionality of collider events, this is typically done by quantifying the agreement of marginalized, one-dimensional projections of the data. The problem with this is that high dimensional, correlated biases might not manifest in low dimensional observables. It is not necessarily possible to construct a single `summary' observable or set of observables which contains information about the bias. Even if there is, it is unclear how to know what they are.\footnote{One way to choose optimal observables using information theory was explored in \cite{lin_ai-driven_2026}.} Studying multidimensional sets of observables to understand these high dimensional correlations also becomes impossible very quickly due to the curse of dimensionality. Additionally, the analysis of binned observables is sensitive to the choice of binning. A perceived agreement between two 1D observables could be an artifact of any of these decisions even if the underlying datasets are incompatible. Due to these issues, we have applied the Cross Section Mover's Distance ($\Sigma$MD)~\cite{komiske_hidden_2020} as an unbinned, full-phase-space method to quantify the bias between datasets \cite{doherty_optimal-transport-based_2026}. This method is rooted in optimal transport (OT) theory as the distance between the original and target dataset. 

\subsection{The Energy Mover's Distance}
The Energy Mover's Distance (EMD) quantifies the work required to transform one collider event into another and is a metric on the space of collider events. The EMD is a true metric that satisfies all of the metric axioms of positivity, symmetry, identity of indiscernibles, and triangle inequality \cite{komiske_metric_2019, komiske_hidden_2020, ba_shaper_2023}. An event with $N$ particles can be represented by its energy flow: 

\begin{equation}
    \mathcal{E}(\hat{n})=\sum_{i=1}^NE_i\delta(\hat{n}-\hat{n}_i)
\end{equation}
where $E_i,\hat{n}_i$ are the energy and angular position of particle $i$ respectively. For proton collisions at the Large Hadron Collider (LHC), particle energies are replaced with their transverse momentum ($p_T$) and their angular positions are parameterized by the pseudorapidity ($\eta$) and azimuthal angle ($\phi$). The EMD between two events $\mathcal{E}, \mathcal{E}'$ takes the following form:

\begin{equation}
    \text{EMD}_{\beta,R}(\mathcal{E},\mathcal{E}')=\min_{f_{ij}>0}\sum_{i=1}^N\sum_{j=1}^{N'}f_{ij}\left(\frac{\theta_{ij}}{R}\right)^\beta + \left|E_\text{tot}-E'_\text{tot}\right|
    \label{eq:EMD}
\end{equation}
where $f_{ij}$ is the transport plan quantifying the energy movement from particle $i\in\mathcal{E}$ to $j\in\mathcal{E}'$ subject to the following constraints:

\begin{equation}
    f_{ij}\geq0,\quad\sum_{j=1}^{N'}f_{ij}\leq E_i, \quad\sum_{i=1}^{N}f_{ij}\leq E_j', \quad\sum_{i=1}^{N}\sum_{j=1}^{N'}f_{ij}=\mathrm{min}(E_\mathrm{tot},E'_\mathrm{tot}).
\end{equation}
The first term in Eq.~\ref{eq:EMD} quantifies the cost of transforming one collider event into another and the second term accounts for events with different total energies. There are three user-specified inputs in this metric. First is the ground metric $\theta_{ij}$ between particles on the detector, which we choose to be Euclidean in the $\eta\phi$ plane with periodic $\phi$. Second is the parameter $R$, which controls the relative importance between the first and second terms of the EMD. Third, the angular exponent $\beta>0$ controls the sensitivity to energy transport at different scales. To obey the generalized triangle inequality, the first term in Eq.~\ref{eq:EMD} should be taken to the $\frac{1}{\beta}$ power for $\beta\geq 1$. One of the benefits of the EMD is that it is infrared and collinear (IRC) safe, meaning the distance is robust to soft emissions and collinear splittings. This is significant because observables constructed from EMDs can be perturbatively calculable \cite{komiske_hidden_2020, ba_shaper_2023}. 

\subsection{The Cross Section Mover's Distance}
The $\Sigma$MD is another optimal transport based metric which defines a distance between theories \cite{komiske_hidden_2020}. A `theory' as defined in the context of the $\Sigma$MD is made up of a discrete set of collider events and their associated cross sections: $\mathcal{T}=\{\mathcal{E}_i,\sigma_i\}$. A theory $\mathcal{T}$ with $N$ events can be represented as a `cross section flow':

\begin{equation}
    \mathcal{T}=\sum_{i=1}^N\sigma_i\delta(\mathcal{E}-\mathcal{E}_i).
\end{equation}
The OT distance between theories takes a similar form to the event-level EMD \cite{komiske_hidden_2020}:

\begin{equation}
    \text{$\Sigma$MD}_{\gamma,S}(\mathcal{T},\mathcal{T}')=\min_{F_{ij}>0}\sum_{i=1}^N\sum_{j=1}^{N'}F_{ij}\left(\frac{\Theta_{ij}}{S}\right)^\gamma +\left|\sigma_\text{tot}-\sigma'_\text{tot}\right|.
\end{equation}
Analogous to the EMD, $F_{ij}$ is the transport plan which quantifies the cross section movement between event $i\in\mathcal{T}$ to $j\in\mathcal{T}'$ subject to the following constraints

\begin{equation}
    F_{ij}\geq0,\quad\sum_{j=1}^{N'}F_{ij}\leq \sigma_i, \quad\sum_{i=1}^{N}F_{ij}\leq \sigma_j', \quad\sum_{i=1}^{N}\sum_{j=1}^{N'}F_{ij}=\mathrm{min}(\sigma_\mathrm{tot},\sigma'_\mathrm{tot}).
    \label{constraints}
\end{equation}
The same user-defined features exist in the $\Sigma$MD as the EMD. The ground metric $\Theta_{ij}$ determines the distance between events, which we choose to just be the event-level EMD: $\Theta_{ij}=\text{EMD}_{\beta,R}(\mathcal{E}_i,\mathcal{E}_j')$. The parameters $S,\gamma$ fulfill the same role as $R, \beta$ in the event-level EMD. It is important to note that the $\Sigma$MD is independent of any single or multidimensional projections, and is constructed out of low level particle features. For theories that are discretely sampled from an unknown underlying distribution, if both theories are sampled from the same distribution, then the $\Sigma$MD between them will approach zero in the infinite sample limit. As the underlying distributions diverge, the $\Sigma$MD grows \cite{komiske_hidden_2020}. 

\subsection{Summary of Related Works}
Several methods have been developed to determine compatibility between high dimensional datasets, such as global statistical tests or training classifiers to determine local fluctuations \cite{diefenbacher_generative_2026}, but these methods are insufficient for our task. This is because negative weights are problematic for classifier based methods and global statistical tests make use of high level features. These methods have been used extensively to test the performance of generative models. The generative model will produce a different dataset than the numerical method, then both are compared \cite{alanazi_simulation_2021, buhmann_getting_2021, butter_how_2019, erdmann_generating_2018, erdmann_precise_2019, hashemi_lhc_2019, musella_fast_2018, otten_event_2021, paganini_calogan_2018, sipio_dijetgan_2019}. We aim to understand the compatibility for a single set of events with different sets of weights.

Global test statistics are typically forms of integral probability metrics (IPMs), which measure the distance between probability distributions by seeing how they behave under a class of functions. IPMs include methods such as sliced Wasserstein distance, maximum mean discrepancy, and others \cite{kansal_evaluating_2023, grossi_refereeing_2024}. These methods typically rely on choosing a set of high level features that sufficiently represents the event, such as energy flow polynomials (EFPs) or learned features from a neural network, then computing a distance between them. Extending this to low level features is problematic because these methods can be sensitive to soft radiation \cite{grossi_refereeing_2024}. The $\Sigma$MD is a global figure of merit which uses only low level particle features and is immune to this infrared effect faced by IPM through IRC safety.

Classifier based methods try to see if a trained classifier can determine which dataset an individual event came from: $x\sim P_1(x)$ or $x\sim P_2(x)$ \cite{grossi_comparing_2025, cappelli_learning_2025}. The likelihood ratio can be computed from the classifier output, which is guaranteed to be the optimal test statistic by the Neyman-Pearson Lemma. The question of how to incorporate weights into the model is unclear with this approach. If weights are treated as a feature, then the classifier will learn to distinguish the distribution of weights since the underlying events are identical. Incorporating weights in the loss function can lead to numerical instabilities due to the negative weights.

\section{Bias in Negative Weight Mitigation}
As a concrete application of the $\Sigma$MD, we apply it to quantify the bias incurred during negative weight mitigation \cite{doherty_optimal-transport-based_2026}. Negatively weighted events are created at next to leading order (NLO) in Monte Carlo generators and greatly dilute the statistical power of the dataset, in addition to increasing the overall computational burden. Several methods have been developed to deal with this issue \cite{beekveld_logarithmically-accurate_2025, doherty_optimal-transport-based_2026, andersen_efficient_2023, andersen_unbiased_2022, andersen_cell_2024, andersen_precision_2026, jadach_matching_2015, jansen_sampling_2025, nachman_neural_2020, olsson_resampling_2020, shyamsundar_arcane_2025}, including different reweighting schemes, but bias is still quantified by studying a set of
observables. Using the $\Sigma$MD, we can rephrase the task of quantifying bias into a purely geometric problem. The distance between the original and reweighted theories is itself the measure of bias and the quantity one wishes to minimize through their proposed reweighting method. 

\subsection{Negative Weights in the $\Sigma$MD}
\label{negweight}
To apply the $\Sigma$MD to quantify bias in negative weight mitigation, one subtle point must be addressed. So far, we have assumed all events are positively weighted. This is a necessary condition because the presence of negative weighted events spoils the first constraint of Eq.~\ref{constraints} and the $\Sigma$MD becomes ill-defined. Computing OT distances between signed measures has been studied in \cite{piccoli_wasserstein_2019} and we adopt their methods. We present the most salient points below.

Suppose we have two signed measures $\mu,\nu$. These can be written as the difference between positive definite measures: $\mu=\mu^+-\mu^-,\nu=\nu^+-\nu^-$. The $^{+,-}$ respectively signify the positive and negative portions of the signed measure. Since there can be transport between any of the measures, there are three interactions which should be encoded in the distance. 
\begin{quote}
\begin{itemize}
    \item Movement of weight between the positive measures ($\mu^+\leftrightarrow\nu^+$)
    \item Movement of weight between the negative measures ($\mu^-\leftrightarrow\nu^-$)
    \item Cancellation of weight (auto-annihilation) between positive and negative measures within a signed measure ($\mu^+\leftrightarrow\mu^-$ $\text{or}$ $\nu^+\leftrightarrow\nu^-$)
\end{itemize}
\end{quote}

A simple definition which has all desired interactions takes the following form:

\begin{equation}
    W_p(\mu,\nu)=W_p(\mu^++\nu^-,\mu^-+\nu^+)
    \label{qpdf_ot}
\end{equation}
where the $p$-Wasserstein distance $W_p(\mu,\nu)$ is the OT distance between $\mu,\nu$ with angular exponent $p$. In the above equation, portions of $\mu^+$ are transported to $\mu^-$ (third point) and $\nu^+$ (first point). Similarly, the transport of $\nu^-$ satisfies the second and third points. However, it is important to note that Eq.~\ref{qpdf_ot} only satisfies the triangle inequality for $p=1$. 

As presented, Eq. ~\ref{qpdf_ot} becomes computationally taxing, especially when the dataset size is large. Luckily, we can take advantage of another property of 1-Wasserstein distances:

\begin{equation}
    W_1(\mu,\nu)=W_1(\mu+\eta,\nu+\eta)
    \label{qpf_ot2}
\end{equation}
where $\eta$ is any arbitrary measure. If we choose $\eta$ to be a constant $\eta=K\geq \left|\min(\mu,\nu)\right|$ then by construction, $\mu,\nu\geq 0$ and we can perform standard OT between $\mu,\nu$. This is independent of the choice of $K$ since Eq.~\ref{qpf_ot2} is insensitive to the choice of $\eta$. It is important to note that Eq.~\ref{qpdf_ot} and Eq.~\ref{qpf_ot2} are equivalent for $p=1$ but Eq.~\ref{qpf_ot2} is implemented in practice because it is less computationally expensive.\footnote{Implementation details for the $\Sigma$MD can be found in \cite{otcres_v1.0.0}.} However, this equivalence holds only when both theories share the same support. If this is not the case, supports can be matched by adding `ghost' events with weights of zero until both theories have the same set of events \cite{doherty_optimal-transport-based_2026}.

\subsection{Results}
We compute the $\Sigma$MD between two theories $\mathcal{T}=\{\mathcal{E}_i,\sigma_i+K\}$ and $\mathcal{T}'=\{\mathcal{E},\sigma_i^\text{RW}+K\}$, where both theories have the same events but differ in the weights. $\sigma_i$ are the original weights from the generator, $\sigma_i^\text{RW}$ are the weights after a reweighting scheme, and $K$ is the constant discussed in Section ~\ref{negweight}. The $\Sigma$MD distance quantifies the bias incurred from the reweighting. 

The reweighting method used is known as `cell resampling' \cite{andersen_unbiased_2022, andersen_efficient_2023, andersen_cell_2024, andersen_precision_2026}. This is where weight is locally smeared between events that are near each other. The distance measure between events is a user defined choice and we test several different OT based metrics. We also apply cell resampling at the different stages in the collision process: hard scatter (HS), parton shower (PS), and hadronization (HAD) \cite{doherty_optimal-transport-based_2026}. We conduct reweighting on a sample with $10^5$ events of $Z$+jets and $t\bar{t}$ with $37.6\%$ and $22.8\%$ negative events respectively. Events were generated at NLO using \textsc{MadGraph5\_aMC@NLO}~3.5.6 and \textsc{Pythia}~8.3 \cite{doherty_2026_21284997}. 

\begin{figure}
    \centering
    \sidecaption
    \includegraphics[width=0.7\linewidth]{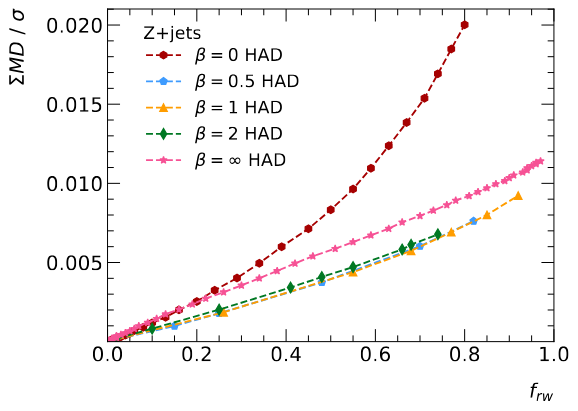}
    \caption{$\Sigma$MD between original and reweighted $Z$+jets sample as a function of reweighted fraction. The cell resampling metrics are the EMD with various values of $\beta$ at the hadronization stage \cite{doherty_optimal-transport-based_2026}.}
    \label{zjets}
\end{figure}

First, we look at five reweighting methods applied to the $Z$+jets sample \cite{doherty_optimal-transport-based_2026}. In Fig.~\ref{zjets}, the $\Sigma$MD rises as a function of the reweighted fraction. This is expected because as more events are reweighted, $\mathcal{T}'$ moves further away from $\mathcal{T}$. We can also see that there is a hierarchy in the bias. The $\beta=0,\infty$ reweighting methods appear to cause the most bias since those $\Sigma$MD distances are the largest. The remaining reweighting schemes have comparable bias \cite{doherty_optimal-transport-based_2026}.

\begin{figure}
    \centering
    \sidecaption
    \includegraphics[width=0.7\linewidth]{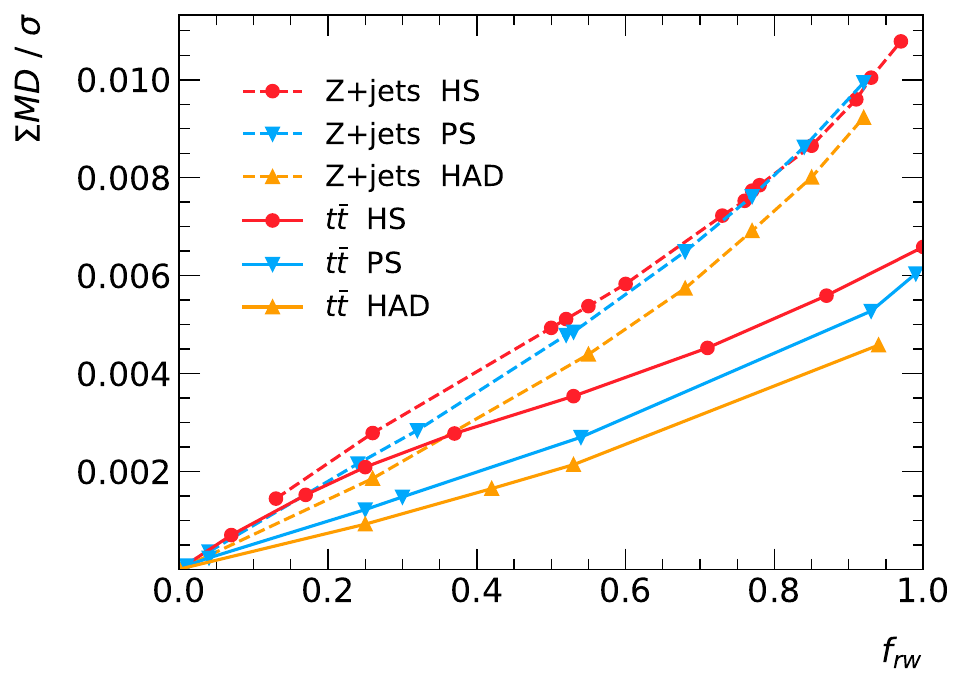}
    \caption{Dimensionless $\Sigma$MD between original and reweighted $Z$+jets and $t\bar{t}$ samples as a function of reweighted fraction. The cell resampling metrics are the $\beta=1$ EMD at different collision stages \cite{doherty_optimal-transport-based_2026}.}
    \label{ttbar}
\end{figure}

Next, we look at how reweighting different samples affects the $\Sigma$MD \cite{doherty_optimal-transport-based_2026}. We apply the same three reweighting methods to the $Z$+jets and $t\bar{t}$ sample then compute the dimensionless $\Sigma$MD by dividing the standard $\Sigma$MD by the total cross section of the process. This is necessary to effectively compare processes with different total cross sections. In Fig.~\ref{ttbar} we see that once again, all the $\Sigma$MDs are increasing as a function of the reweight fraction. There is also a hierarchy of bias with the HAD method causing the least bias across both processes. It is also interesting to note that the $t\bar{t}$ events have a lower $\Sigma$MD than the $Z$+jets events for a given reweight fraction. This can be explained by the fact that the $t\bar{t}$ sample has less total negative events. For a given reweight fraction, less $t\bar{t}$ events are affected which naturally results in less bias compared to the $Z$+jets sample \cite{doherty_optimal-transport-based_2026}. 

\section{Conclusion}
We have introduced the $\Sigma$MD as a novel global test statistic in quantifying bias from negative weight mitigation by reframing the bias as a distance between theories. The $\Sigma$MD is computed entirely using low level particle features and is insensitive to soft radiation. We apply this to datasets reweighted with the cell resampling method with various OT based metrics. As expected, the bias computed from the $\Sigma$MD increases as more negative events are reweighted. We are also able to determine which OT based metrics cause the least bias as well as how the $\Sigma$MD depends on the underlying physics process and collision stage. 

Several approaches to negative weight mitigation have been developed in addition to cell resampling. The $\Sigma$MD presents a common way to measure performance across methods. More broadly, reweighting is performed in many situations such as unfolding or applying detector corrections, and the $\Sigma$MD can find success in comparing the reweighted sample with the original. The $\Sigma$MD is not restricted to cases where only the weights are changing. For example, when comparing generative models to their numerical counterparts, the $\Sigma$MD can be a complementary, global figure of merit.

\section*{Acknowledgments}

This material is based upon work supported by the U.S. Department of Energy, Office of Science, Office of High Energy Physics under Award Number DE-SC0026285 and is supported by the National Science Foundation under Cooperative Agreement PHY-2019786 (The NSF AI Institute for Artificial Intelligence and Fundamental Interactions, \href{http://iaifi.org/}{http://iaifi.org/}).

\printbibliography

\end{document}